\documentclass[notitlepage,aps,pra,twocolumn,groupaddress]{revtex4-1}

\usepackage{graphicx}
\usepackage[colorlinks=true, allcolors=blue]{hyperref}

\usepackage{stmaryrd}
\usepackage{amssymb,amsmath,amsthm,amsfonts,amsbsy,tabularx}
\usepackage{bm,bibunits,color,chngcntr,epsfig,epstopdf,graphicx,dsfont}
\usepackage{hyperref,lipsum,,makecell,mathrsfs,rotating,setspace}
\usepackage[english]{babel}
\usepackage[normalem]{ulem}

\newcommand{\be}{\begin{equation}}
\newcommand{\ee}{\end{equation}}
\newcommand{\bea}{\begin{eqnarray}}
\newcommand{\eea}{\end{eqnarray}}
\newcommand{\ket}{\rangle}
\newcommand{\bra}{\langle}

\newcommand{\I}{\mathds{1}}

\newcommand{\ba}{\begin{align}}
\newcommand{\ea}{\end{align}}
\newcommand{\Tr}{\text{tr}}

\def\C#1{\mathcal #1}
\def\B#1{\mathbb #1}

\begin{document}

\title{Quantum Sequential Circuits}
\author{Dong-Sheng Wang}\email{wds@itp.ac.cn}
\affiliation{Institute of Theoretical Physics, Chinese Academy of Sciences, Beijing 100190, China \\
School of Physical Sciences, University of Chinese Academy of Sciences, Beijing 100049, China}

\newtheorem{theorem}{Theorem}
\newtheorem{prop}[theorem]{Proposition}
\newtheorem{corollary}[theorem]{Corollary}
\newtheorem{open problem}[theorem]{Open Problem}
\newtheorem{conjecture}[theorem]{Conjecture}
\newtheorem{definition}{Definition}
\newtheorem{remark}{Remark}
\newtheorem{example}{Example}
\newtheorem{task}{Task}
\newtheorem{protocol}{Protocol}
\newcolumntype{Y}{>{\small\raggedright\arraybackslash}X}

\begin{abstract}
This work introduces and characterizes quantum sequential circuits (QSCs) 
as a hardware-oriented paradigm for quantum computing, 
built upon a novel foundational element termed the quantum transistor. 
Unlike conventional qubit-based architectures, 
QSCs employ symmetry-protected topological junctions 
where quantum gates are encoded as Choi states via channel-state duality 
and activated through bulk measurements, 
utilizing ebits to realize the functional analog of feedback loops in classical sequential circuits. 
This framework establishes a universal model for quantum computation 
that inherently incorporates memory and temporal sequencing, 
complementing existing combinational quantum circuit model. 
Our work advances the conceptual bridge towards a quantum von Neumann architecture, 
underscoring the potential of hybrid and modular design principles 
for the development of large-scale, integrated quantum information processors.
\end{abstract}
\date{\today}

\maketitle

\begin{spacing}{1.2}

\section{Introduction}

A computer is a complex and controllable physical system 
built upon the von Neumann architecture 
with hierarchical hardware and software layers~\cite{HH13}. 
Its mathematical foundation relies on representing information with bits, 
which are processed through arrays of logic gates. 
The core of the system consists of large-scale integrated circuits, 
composed of analog, combinational, and sequential circuits. 
These enable diverse functionalities such as 
computation, control, transmission, and storage. 
Beyond basic passive components like resistors, capacitors, and wires, 
the transistor stands as the most critical element~\cite{SN07}, 
performing essential roles including switching, gating, storage, and amplification
as a foundational, scalable, and standardized active building block.

In recent years, research has advanced toward the development of quantum computers~\cite{NC00,LJL+10,Pre18,LSS+26}. 
Unlike classical architectures based on transistors and logic gates, 
quantum computing employs qubit architectures implemented with physical systems 
such as trapped ions, neutral atoms, or superconducting circuits~\cite{NC00}. 
Quantum gates and measurements are executed in real time using 
lasers or electromagnetic waves. 
Although certain optical devices can act as quantum gates, 
they are not universal~\cite{KMN+07}. 
Due to the inherent instability of qubits, 
error correction via encoding is indispensable and 
requires rapid classical computation for error analysis~\cite{LB13,BE21}. 
Consequently, classical control instructions are necessary to manage qubit operations, 
and quantum chips are viewed as components 
that can be integrated into existing computational frameworks.
However, it lacks a native hardware element analogous to the 
transistor, limiting the potential complexity and scalability of quantum system architectures.


Digital electronic circuits comprise both combinational and sequential circuits, 
with the latter permitting feedback loops. 
Combinational circuits are typically designed for a fixed input size, 
and altering this size can significantly change the circuit design. 
In contrast, sequential circuits process inputs 
along the time dimension without modifying their spatial structure, 
hence the term ``sequential." 
Feedback loops are essential for constructing memory elements, including flip-flops and registers, 
which retain past input states and compute outputs based on both current inputs and historical data. 
This enables critical functionalities such as clock-driven synchronization, 
temporary storage or caching, iterative operations, 
and algorithms that accommodate indefinite input sizes.


In this work, we investigate quantum circuit design from a hardware perspective. 
The quantum circuit model, analogous to the Boolean circuit model, 
is well established in quantum computing~\cite{NC00}. 
These circuits are fundamentally combinational in nature. 
A longstanding open question is whether and 
how one can define quantum sequential circuits. 
Directly looping a quantum output state back to the same gate is difficult to conceive
as it is unclear what are the quantum analog of loops.

A central concept in sequential circuits is memory, 
which differs from simple bits. 
Memory takes various forms, such as internal and external storage. 
In contemporary quantum computing, qubits serve as the basic building blocks, 
with no strict distinction among qubits, quantum memory, and quantum registers~\cite{Mem}. 
Recent studies on quantum von Neumann architectures have revealed the unique role of quantum memory, 
particularly for storing quantum programs, 
and have identified ebits as the fundamental elements of quantum memory~\cite{W24rev}. 
This is grounded in a key principle of quantum mechanics known as channel-state duality, 
which states that any quantum channel, including unitary gates, 
can be equivalently represented as a state, commonly referred to as a Choi state~\cite{Jam72,Cho75}.
This duality suggests that a quantum gate can exist as a static resource state, 
inspiring the exploration of a new type of quantum hardware element 
that physically embodies this duality and enables the ``storage" and ``on-demand activation" of gates.

In this work, we define quantum sequential circuits (QSCs) 
based on a transistor-style construction of quantum gates. 
We demonstrate that ebits serve as the quantum analog of loops 
in classical sequential circuits. 
In this framework, a gate is encoded in a system with a Hamiltonian and edge modes, 
and gate activation is achieved through measurements on its bulk~\cite{W26}. 
Prior to an input signal, 
the gate exists in the form of a Choi state by virtue of channel-state duality. 
The system possesses symmetry-protected topological (SPT) order~\cite{ZCZ+15}, 
which enables a bulk-edge duality, 
and its states can be represented as a network of ebits, 
up to local operations. 
The role of ebits is to facilitate state transfer or teleportation 
through a gate via measurements. 
Accordingly, a gate must be resettable for repeated use. 
Furthermore, we establish the universality of QSCs 
and describe how to implement quantum algorithms using them, 
as well as hybrid circuits composed of both QSCs and combinational quantum circuits.




\begin{table}[b!]
    \centering
    \begin{tabularx}{0.48\textwidth}{c|Y|Y}\hline
         \normalsize & \normalsize Classical &  \normalsize Quantum \\ \hline
    Bits & Current/Voltage & Edge modes/Bound states \\  \hline 
    Signal     & Current/Voltage & Phase, Amplitude, Particle number etc \\  \hline
    Energy source & Electricity/Battery & Laser/Electricity \\  \hline
    Material & Semiconductor, Metal & Topological materials \\  \hline
    Circuit memory & Electronic loops & Ebits \\  \hline
    Transistors & BJT/FET junctions etc & SPT junctions etc \\  \hline
    Measurements & Passive, Deterministic & Active, Destructive, 
    Probabilistic \\   \hline 
    \end{tabularx}
    \caption{A brief comparison between classical and quantum circuits.}
    \label{tab:architecture}
\end{table}

Our hardware-oriented construction of QSCs 
complements current quantum computing architectures. 
We hereby provide a comprehensive comparison between classical and quantum computational frameworks, 
summarized in Table~\ref{tab:architecture}, 
with an emphasis on hardware implementation. 
The table does not enumerate details of existing platforms, 
and we note that, in principle, other types of QSC constructions may also be possible. 
It highlights that novel quantum materials are required for the quantum case, 
and that lasers are essential for executing gates and measurements while maintaining qubit coherence. 
The complexity exhibited by classical electronic circuits and computers suggests that 
developing diverse types of qubits and quantum circuits, 
along with methods to control and integrate them, will be necessary.



\section{Quantum transistors}

Here we describe the quantum transistors that are studied in details in another work~\cite{W26}.
Despite the name, they are not the direct analog of electronic transistors,
nor the only possibility. 

The central principle of the quantum transistors we defined is that 
a gate is stored as its Choi state following from von Neumann architecture~\cite{W20_choi,W22_qvn},
and the gate operation itself is performed via measurements on 
a symmetry-protected topological (SPT) state, notably, a cluster state~\cite{RB01}
or a valence-bond solid~\cite{AKLT87}, which share the common feature of Haldane phase~\cite{Hal83}; 
namely, there are symmetry-protected edge modes.
These edge modes carry the logical qubits 
and the bulk is measured to induce logical gates. 

\begin{figure}[t!]
    \centering
    \includegraphics[width=0.12\textwidth]{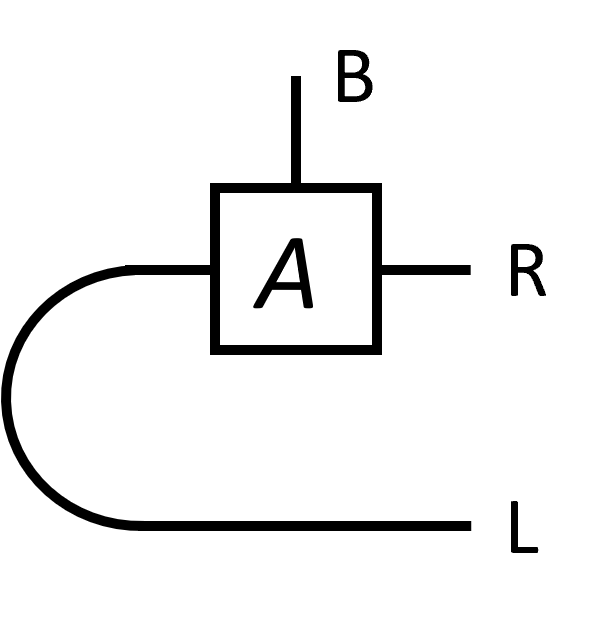}
    \caption{A representation of the quantum transistor we defined in the form of Choi state.
    The curve on the left represents ebits.}
    \label{fig:qtrans}
\end{figure}

It is known that a chain of cluster state can be used to realize any qubit gate
based on gate teleportation. 
The state has $Z_2\times Z_2$ symmetry which defines a wire basis for its representation. 
Such a basis is protected by the symmetry. 
If the measurement is restricted to be in the wire basis,
it can only realize state transfer, i.e., the identity gate~\cite{ESB+12}.
To see this, we start from the matrix-product state (MPS) representation of a state
\be |\psi\ket = \sum_{i} \Tr (E A_i) |i\ket \ee 
for $i:= i_1 i_2 \cdots i_N$, $A_i= A_{i_N} \cdots A_{i_1}$, and $E$ as an edge operator~\cite{PVW+07}.
The logical data is in the bond space acted upon by these $A_i$ and $E$. 
For open boundary condition, $E=|l\ket \bra r|$ and $|l\ket$ is viewed as 
the initial state of the edge mode, and $|r\ket$ is the final projection. 
In order to avoid them, 
additional qubits coupled to the left mode and others to the right mode can be used to form ebits,
and the whole state becomes
\be |\Psi\ket = \sum_{ij} |j\ket_\text{L}  A_i |j\ket_\text{R} |i\ket_\text{B}, \ee 
and we use the subscript L, R, B to represents the left mode, right mode, and the bulk.
Note we just call the additional qubits on the left as the left mode since 
its role is to make the ``virtual'' mode physical. 
See figure~\ref{fig:qtrans}, and we can see it has a transistor-style structure: 
the L system as the input, R as the output, and B as the control terminal. 
The input signal is a measurement that prepares an initial state,
e.g., the state $|0\ket$ of a qubit.

For the standard one-dimensional qubit cluster state, 
the wire basis measurement leads to the gate sequence 
\be HZ^{i_{n}} \cdots HZ^{i_2} HZ^{i_1} \ee 
which is a Pauli gate if $n$ is even, 
and a Hadamard gate $H$ (modular a Pauli gate) if $n$ is odd. 
A Pauli gate is often treated as a byproduct that can be easily dealt with.
So we can realize logical identity gate or $H$ gate depending on the system length. 
Similarly, we can also realize the phase gate $S$ by a cluster chain in a rotated basis
and the $CZ$ gate by a two-leg ladder~\cite{W26}. 
These gates are protected by the SPT order of cluster states. 
The $T$ gate is realized via magic-state injection~\cite{BK05}.
Given the universal gate set of $H$, $S$, $T$, $CZ$,
as well as Pauli gates, any other gate can be constructed as a connected gate array.
They are all stored in the form of Choi state,
and for simplicity, we call them as quantum transistors no matter what gate it is.
This is the foundation for the quantum sequential circuits we introduce below.


\section{Circuit elements}

\subsection{Combinational circuits}

We first introduce symbols for elements in usual quantum circuits.
A qubit is represented by a black dot,
a free evolution is represented by an arrow, 
a unitary gate is represented by a box,
and a measurement is represented by a particular meter symbol 
with a double arrow for the classical outcome.
See figure~\ref{fig:circ} and also an example.
The qubits are on the left of the circuit diagram,
and the arrow represents the evolution time direction. 
There cannot be loops in a combinational  circuit meaning that
an output from a gate cannot ``go back'' again to serve as its new input. 

In practice, gates and qubits can be realized in different systems.
There are photonic and solid-state platforms~\cite{LJL+10}.
For photonic platform, qubits are often carried by photons 
which are ``flying'' qubits and generated on-time,
while gates are hardware that occupy the major part of a photonic chip.
However, these gates are not universal, and other schemes have to be employed,
such as fusion-based gates~\cite{BBB21}.
For solid-state platform such as superconducting circuits,
gates are temporal signals generated on-time, 
while qubits are hardware that occupies the major part of a chip.
Lasers are the main devices to generate qubits, perform gates or measurements.

\subsection{Sequential circuits}\label{subsec:seq}

\begin{figure}[t!]
    \centering
    \includegraphics[width=0.25\textwidth]{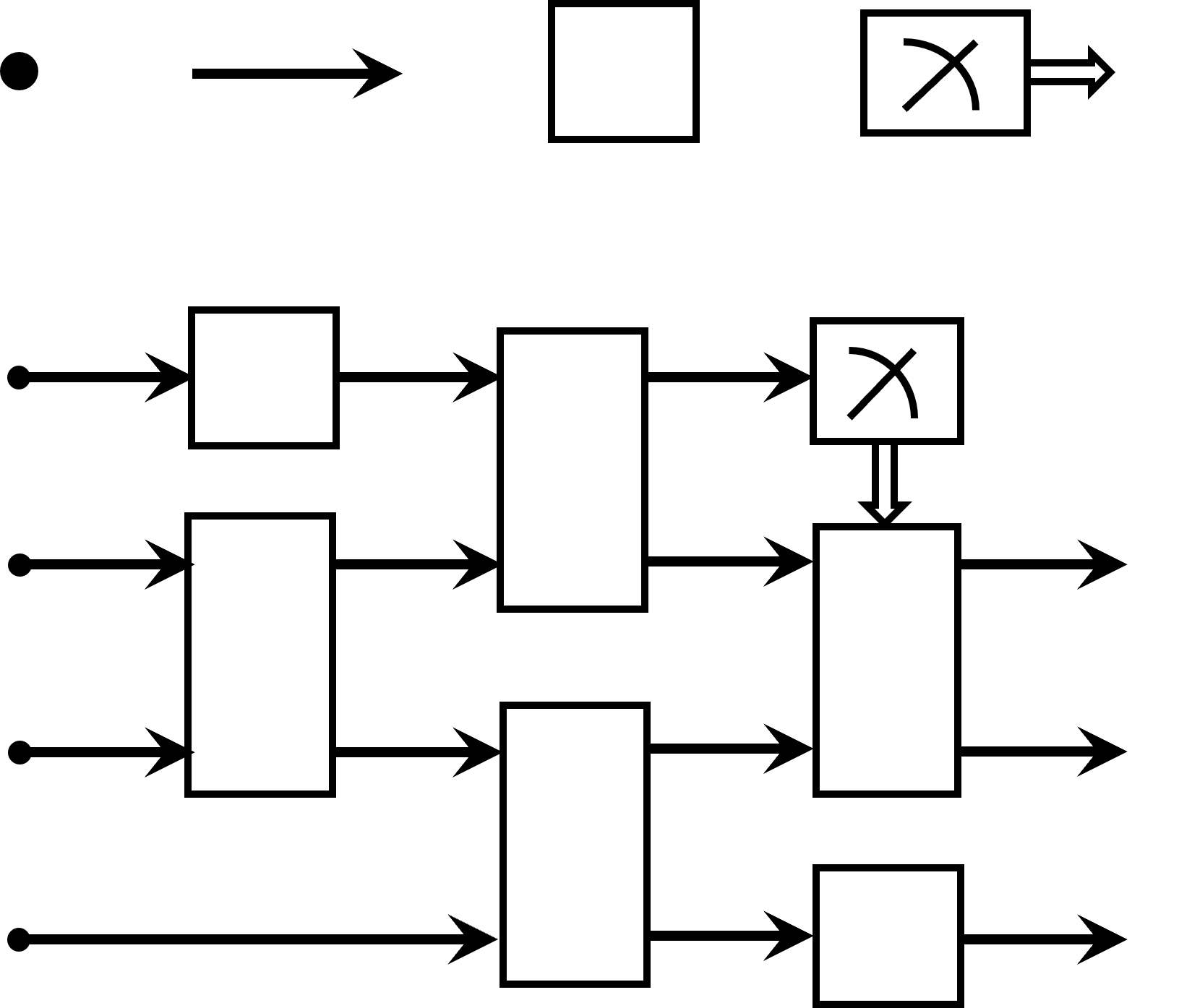}
    \caption{Symbols for elements in combinational  quantum circuits (top)
    and an example circuit (bottom), 
    which also shows a gate conditional on the measurement outcome.
    Note one can also put dots at the end of the circuit to label output qubits.}
    \label{fig:circ}
\end{figure}

\begin{figure}[t!]
    \centering
    \includegraphics[width=0.2\textwidth]{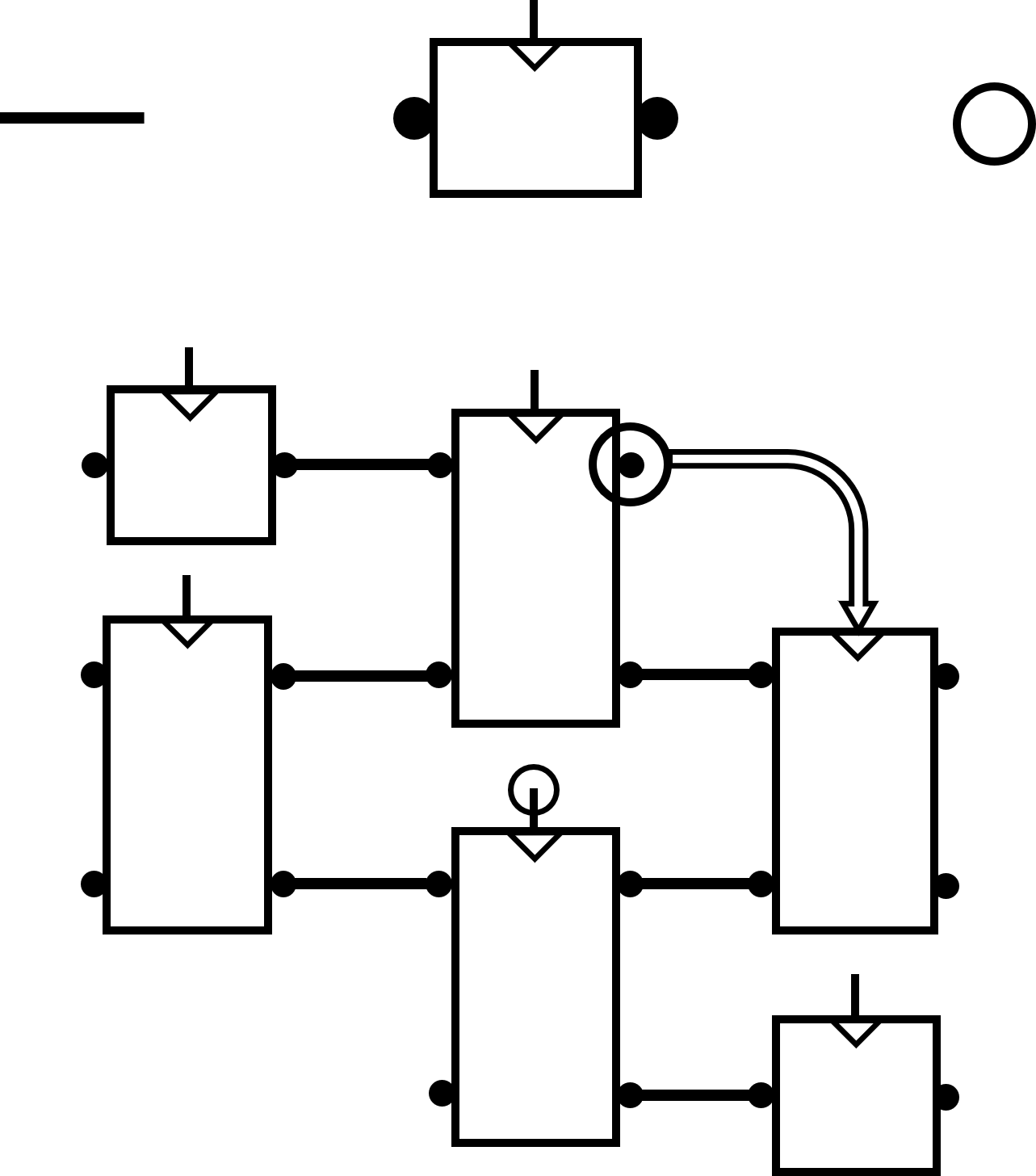}
    \caption{Symbols for elements in sequential quantum circuits (top)
    and the same circuit as in figure~\ref{fig:circ} (bottom).
    A measurement signal on a gate is also shown.}
    \label{fig:circ2}
\end{figure}

We now introduce the elements for quantum sequential circuits.
An ebit (Bell state) is represented by a short line or even curved line,
a stored gate is represented by a box with two black dots on its left and right sides,
and also a control symbol at the top,
and a measurement signal is represented by a circle.
See figure~\ref{fig:circ2} and the same circuit from 
figure~\ref{fig:circ}.
Note that there are external qubits carried by ``quantum dots'' coupled to the bulk 
to form physical edge modes, hence the two dots for its symbol.
Before the measurement signal, there is no initialized qubits and time evolution,
and the whole circuit can be viewed as a stored gate in the form of Choi state.
Each gate is opaque before the bulk measurement,
and becomes transparent after it 
and the input is mapped to the output.

\begin{figure}[b!]
    \centering
    \includegraphics[width=0.25\textwidth]{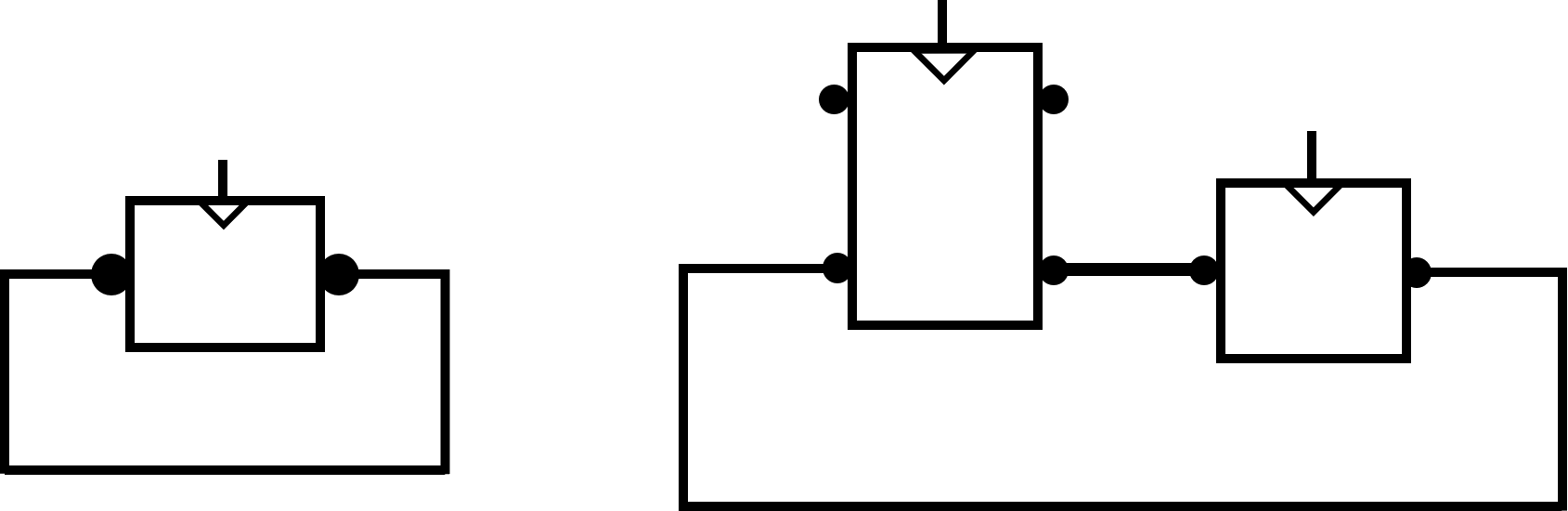}
    \caption{Some simple examples of sequential quantum circuits with loops.
    This type of circuits can be used to realize iterative operations. 
    }
    \label{fig:loop_gate}
\end{figure}

There can be loops for such circuits. 
A loop is actually an ebit (or a product of them), 
and this forms the periodic boundary condition for the bulk system.
Some examples are shown in figure~\ref{fig:loop_gate}.
Such an ebit can also be augmented by a chain of cluster state or 
other states of the same nature, and the goal is to teleport the output 
back to the input edge mode. 
This does not violate causality since the time order is set by the measurements.
As measurement is destructive, especially the bulk measurement,
a gate is one-time and has to be refreshed for the next use.
As long as the system Hamiltonian is controllable, 
the gate can be refreshed~\cite{W26}. 

To use loops to realize a reusable gate,
a sequence of measurements and refreshments can be used.
An elementary procedure is:
\begin{spacing}{1.0}
\begin{enumerate}
    \item Measure the left mode to prepare input;
    \item Measure the bulk to execute the gate, and the data is transferred to the right mode;
    \item Refresh the left mode and the bulk;
    \item Establish ebit between the left and right modes;
    \item Measure the right mode to transfer data to the left mode;
    \item Refresh the right mode;
    \item Back to the 2nd step and repeat.
\end{enumerate}
\end{spacing}

The procedure above is a sequence of gate teleportation~\cite{GC99,ZLC00},
and it can also be realized by a usual combinational  circuit. 
This points to the central feature of sequential circuits, 
which is to use the space-time conversion to save space cost using refreshable gates. 
As for electronic circuits, 
we expect it can benefit future quantum chip design,
and we will show its usage for quantum circuits and algorithms in the following sections.

\subsection{Synchronization}\label{sec:syn}

Timing is a central part in the design of electronic circuits.
Proper sequential circuits are synchronized.
One might expect to use superposition of timing for the quantum case. 
In the original setting of quantum Turing machine, 
this has been analyzed and shown to add complexity to the computing process~\cite{BV97}.
Here we use classical timing or clock, 
and will study quantum timing in Sec.~\ref{sec:qst}.
As the control signals are for measurements,
it is natural to use classical clocks to control the order of gates.
Proper timing is relevant for the design of various sequential elements,
such as latch and flip-flop.
Some of these designs can be carried over to the quantum case 
as our timing scheme is classical. 



\subsection{Hybrid circuits}

A circuit can be hybrid, 
containing both combinational  and sequential parts.
For the quantum case, they could be carried by different types of systems,
such as photons, trapped ions, cold atoms, superconducting circuits, quantum dots, 
ground states or excitations of many-body systems.
A hybrid circuit may involve any possible combinations of those systems 
containing qubits, ebits, gates, and transistors which are stored gates.

How to connect a combinational  circuit to a sequential circuit? 
It only needs to establish ebits between the qubits 
in the combinational  circuit and the input edge modes.
If the combinational  circuit follows behind, 
it can be directly realized on the output edge modes,
or by firstly teleporting the edge modes to other qubits,
and then running the circuits on these qubits.

\subsection{Quantum register}

Register is a notion that is different from bit for electronic circuits.
Yet, in quantum computing 
a quantum register often refers to a collection of qubits.
In our model, a quantum register is a collection of quantum transistors
with their input modes carrying an entangled data state $|\psi\ket$.
Normally, the quantum transistor shall only execute the identity gate;
however, we also allow other gates (such as $H$ and $S$ gates) 
for the construction of quantum registers. 
A type of register is known as \emph{shift register}, 
which can shift the data across its transistors.
The shift function can be easily realized by quantum teleportation, 
therefore, it is straightforward to define quantum shift register
by connecting transistors in series.

\subsection{Quantum finite state machine}

\begin{figure}[t!]
    \centering
    \includegraphics[width=0.33\textwidth]{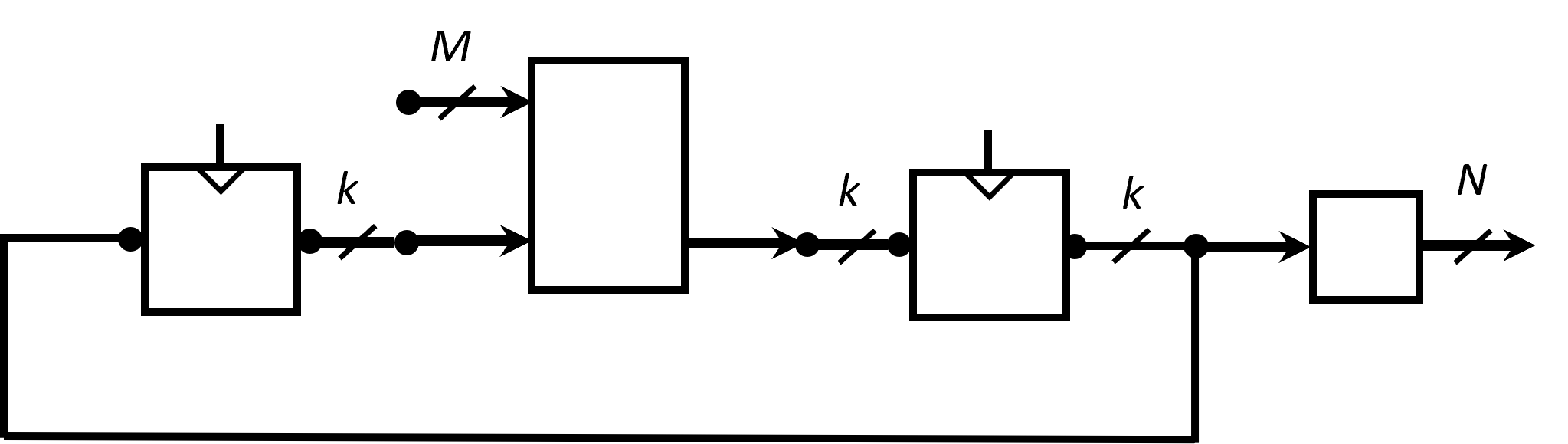}
    \caption{Quantum finite state machine which is a type of hybrid circuit. 
    }
    \label{fig:qfsm}
\end{figure}

Finite state machine is a type of synchronized hybrid circuits.
We find that it carries over to the quantum case,
and its basic structure is shown in figure~\ref{fig:qfsm}.
It actually has a form of MPS,
with the registers acting on its bond space,
and it has been shown that MPS is the underlying mathematical language for 
quantum finite state machine and the local model of quantum Turing machine~\cite{W20_T}. 
Recall that MPS does not necessarily refer to one-dimensional systems;
instead, it can describe systems of arbitrary geometry.

The figure shows a circuit with $M$-qubit input and $N$-qubit output,
and with $k$-qubit quantum registers.
It is of the so-called Moore-style, 
and the Mealy-style is also possible by allowing some input 
directly go into the final combinational  circuit.

\subsection{Quantum pipeline}

Pipeline circuitry breaks down complex operations into simpler stages, 
allowing multiple instructions to be processed simultaneously, 
which significantly improves overall throughput.
It saves time to execute an algorithm without a major hardware cost.

A quantum scheme that naturally serves as a quantum pipeline protocol is 
the stabilizer formalism~\cite{Got98,NC00}.
A circuit can be divided into blocks of $T$ gates and Clifford gates.
The gates in a Clifford circuit can be run in parallel, namely, 
all measurements can be done in parallel for the transistors in a Clifford circuit.
Each measurement will generate Pauli byproduct,
and as Clifford gates preserve the Pauli group,
all byproduct will gather together to form a net Pauli byproduct 
that can be corrected before the $T$ gates. 
This feature has been observed in the context of the standard circuit model
and also MBQC~\cite{RBB03}. 
Beyond this, there could also be other quantum pipeline designs that deserve separate study.

\section{Quantum algorithmic modules}

In this section, we analyze some common quantum algorithms and their primitives,
and show how to use QSCs to deal with iteration, temporal storage, 
or jump instructions.

\subsection{Quantum control and multiplexer}\label{sec:qmux}

\begin{figure}[t!]
    \centering
    \includegraphics[width=0.45\textwidth]{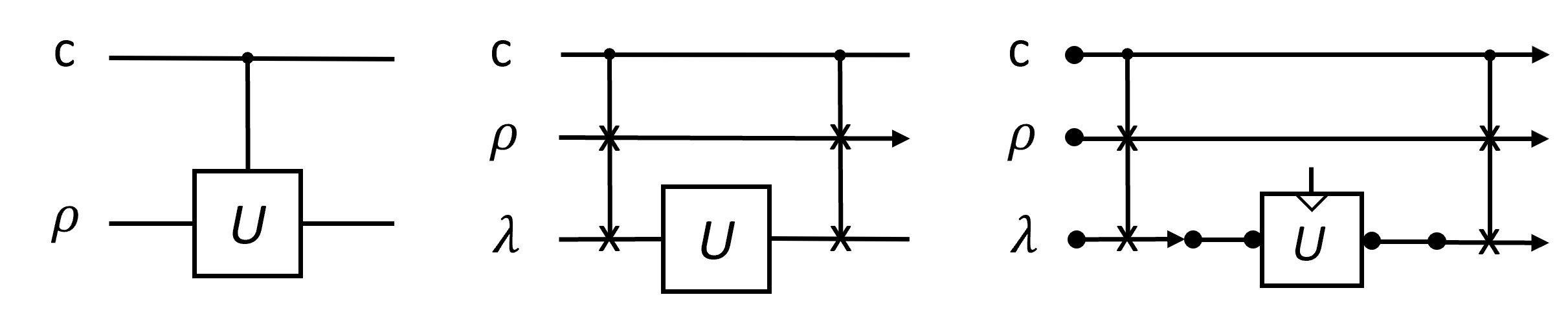}
    \caption{Quantum control operation and its realization with quantum transistor
    in a hybrid circuit. 
    }
    \label{fig:qcontrol}
\end{figure}

Quantum control operations are common in quantum circuits and algorithms. 
If the control is followed by a measurement, 
it can be replaced by a classical control.
If the target gate is unknown,
this cannot be directly realized; instead, 
the circuit shown in figure~\ref{fig:qcontrol} is needed~\cite{AFC14}. 
Here, if the target gate is stored,
it can be realized as the bottom panel shows. 
In the circuit, c is the control qubit, 
$|\lambda\ket$ is an eigenstate of the gate $U$, 
and there are two controlled-swap gates. 
Note that we choose to use hybrid circuit to realize this.
One can also use a purely sequential circuit that will involve 
more transistors to store all the qubits and realize the controlled-swap gates
by decomposing them into elementary gates. 

This can be extended to multiple target gates.
Denote a single controlled gate as $\wedge_U=\text{diag}(\I, U)$, 
then a quantum multiplexer (QMUX) is diag($U_1, U_2, \dots, U_n$),
and it can be realized in a similar way. 
The QMUX module is widely used in quantum algorithms,
such as the quantum phase estimation~\cite{KSV02}, 
Shor algorithm~\cite{Sho94}, swap test, DQC1~\cite{KL98}, and 
linear combination of unitary gates~\cite{CW12}, etc.

\subsection{Quantum Fourier transformation}

The quantum Fourier transformation $F$ is also a primitive in quantum algorithms~\cite{NC00}. 
The Hadamard gate $H$ is the qubit version of it.
On $n$-qubits, it can be realized by a circuit of $O(n^2)$ gates, 
and it uses the controlled-phase gates with the phase rotation gates 
$R_k=\text{diag}(1,e^{i2\pi/2^k})$ for $k\in [2,n]$. 
The phase depends on $n$ exponentially.
In many cases, the inverse $F^\dagger$ is at the circuit end,
so the control can be replaced by classical control,
and all gates become single-qubit operations.

There is also an analog interpretation of $F$,
which is to view it as the basis transformation between a coordinate and its momentum basis.
If we view qubits as spins,
the measurement can be treated as the measure of the dispersion relation $E(p)$
for the height signifying the probability of a momentum value $p$.
In quantum magnetism, this can be realized by neutron scattering experiment~\cite{Gia04}.
We will see below this point of view can benefit our understanding of some quantum algorithms.

\subsection{Transpose and inverse operations}

It is straightforward to realize the transpose of an operator $A$
according to the basic property of ebit $(A \otimes \I)|\omega\ket=(\I \otimes A^t)|\omega\ket$,
and this extends to any gate or channel.
This is realized by exchanging the input and output modes of a transistor, i.e.,
run a transistor backwards.
An interesting fact to note is that gates $H$, $S$, $CZ$, 
and also $T$ are all symmetric matrices,
so for each gate any one of their edge modes can be chosen as the input mode.

Running a combinational  circuit $U$ backwards realizes its inverse $U^\dagger$
instead of its transpose $U^t$.
However, if $U$ is unknown, it is not easy to realize $U^\dagger$.
If $U$ is given as a network of transistors, 
running it backwards directly realizes $U^t$.
To get $U^\dagger$, one only needs to further change all gates $S$ and $T$ to their conjugates.
For $S$ gate, this can be achieved by appending a Pauli $Z$ operator, 
and for $T$ gate realized via the magic-state injection scheme, 
this can be achieved by simply modifying the classical feedforward and byproduct correction. 
Therefore, only partial knowledge of $U$ is enough to realize $U^\dagger$.

The partial transpose operation can also be easily realized. 
This can be explored to simplify hardware design.
For instance, to prepare a MPS by a sequential circuit with a unitary $U_n$
for each tensor $A_n$, the transpose of a multipartite gate $U_n$
can be used to generate a tensor or its transpose. 
This can also be viewed as a change of boundary condition via 
complex loops of ebits~\cite{W22_qvn}. 
Another notable instance is the swap gate whose partial transpose is 
nothing but the ebit projector $|\omega\ket\bra \omega|$.
Therefore, the swap-gate module can be used as projectors,
or even as generators of periodic boundary condition. 

\subsection{Quantum search and amplitude amplification}

Grover's search algorithm is a novel start of iterative quantum algorithms~\cite{Gro96}.
It has been extended by quantum walk, amplitude amplification,
singular-value transformation and others~\cite{MRTC21}.
It is natural to realize these algorithms as sequential circuits with loops. 

In quantum amplitude amplification (QAA)~\cite{BHM02}, the goal is to drive a parameter $p\in (0,1)$
close to 1 by an iterative gate $Q$ so that a unitary $U$ is realized with 
\be Q^n A |0\ket |\psi\ket \approx  |0\ket U |\psi\ket  \ee 
for $A |0\ket |\psi\ket = \sqrt{p} |0\ket U |\psi\ket + \sqrt{1-p} |\Phi\ket$, 
and $\bra 0|\Phi\ket=0$.
The ``walk'' operator $Q$ can be stored as a transistor,
and its iteration can be realized via the method described in section~\ref{subsec:seq}.
The initial state $|0\ket |\psi\ket$ and the gate $A$ can be realized 
as in combinational  circuit or stored depending on the problem.
The inverse $A^\dagger$ may be needed to realize $Q$,
and it can be realized according to the last section. 

Traditional electronic transistors have the unique function of signal amplification.
This does not have a direct quantum analog, though.
Here, the QAA is rather an \emph{algorithmic} amplification,
and the quantum `signal' is a real amplitude parameter. 
The algorithm is iterative and unitary (i.e. reversible),
while the electronic amplification is apparently irreversible. 

It is possible to induce non-unitary gates, i.e., channels~\cite{NC00}, 
acting on the edge modes of a transistor, e.g.,
by mixing the bulk measurement outcomes.
Using MPS, we know that for each bulk site the tensor $\{A_{i}\}$ forms a channel.
Whether this can be used to realize quantum amplification,
or other functions, is left for future study.  

The QAA has been generalized by 
the quantum singular value transformation (QSVT)~\cite{GSLW19},
which provides a unique perspective of quantum algorithms and 
has found wide applications especially in quantum machine learning.
Treating the positive signal $p\in (0,1)$ as a singular value,
the QSVT can process a set of signals carried by a matrix 
based on the singular-value decomposition.  
It is a sort of quantum \emph{involution} acting as 
feature extractor by sliding filters across input data to capture spatial or temporal patterns efficiently.
Note there are also other definitions of quantum involution, 
e.g., Ref.~\cite{BCZ23}, or as quantum memory effects 
used for quantum involutional codes studied below in Sec.~\ref{sec:qecc}.

\subsection{Quantum phase estimation}

\begin{figure}[t!]
    \centering
    \includegraphics[width=0.27\textwidth]{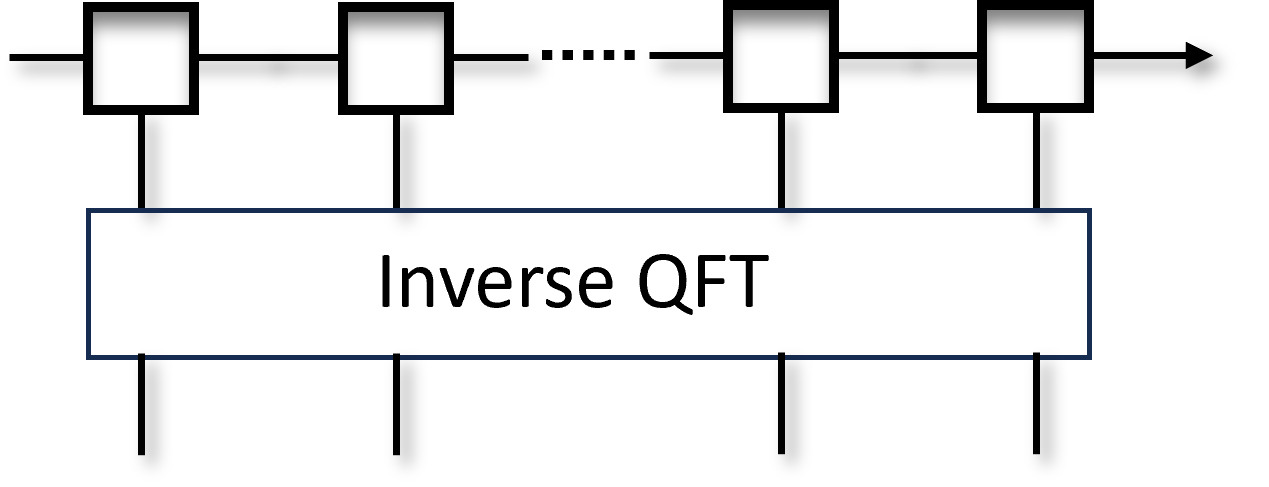}
    \caption{Quantum phase estimation represented as a particular quantum transistor 
    based on the MPS formalism. 
    Note each piece of gate can be realized as a transistor. 
    }
    \label{fig:qpe}
\end{figure}

The quantum phase estimation (QPE) is a central algorithm which estimates the phase $e^{i\theta}$ 
of a gate $U$ on its eigenstate $|\psi\ket$.
It uses a sequence of controlled-gates (the modular exponentiation) and measurement in the 
momentum basis.
Combined with amplitude amplification, 
the QPE can also be extended to amplitude estimation~\cite{BHM02}.

We find that QPE can be viewed as a scheme based on MPS or as a special quantum transistor,
shown in figure~\ref{fig:qpe}. 
Here the MPS is not translation-invariant, and for each site $r$,
the tensor matrices are $A_0=\I$, $A_1=U^{2^{r-1}}$.
For arbitrary $U$, it could be expensive to realize it for exponential many of times,
yet for $U$ with special properties, such as the case in Shor's algorithm
or Hamiltonian evolution where $2^{r-1}$ converts to time, 
or even QAA, 
these gates can be rather easily realized.  

The input edge mode is the initial state $|\psi\ket$ and as it is an eigenstate,
it will also stay as the same state. 
We see that it has the structure of quantum transistor,
and the qubits form the bulk.
The measurement on them is in the Fourier transformed basis,
different from a usual transistor.
Also the measurement outcome carries the final result of the phase value.
In other words, the bulk measurement outcome could be a nontrivial part of an algorithm. 

Here we also remark that recently we introduced the entanglement picture based on MPS formalism~\cite{WXL25}. 
It studies a system by switching from the real physical space to the entanglement space.
What is interesting for the quantum phase estimation is that 
the state $|\psi\ket$ can be viewed as the edge mode of a MPS,
hence we can view this algorithm as an algorithm in the entanglement picture.

\subsection{Quantum gradient descent}

Here we study a primitive method to realize quantum gradient descent,
which is based on the linear combination of unitary operation (LCU) algorithm.
The LCU algorithm has a similar structure with the QPE,
while it allows more general input state $|\psi\ket$ and its function is different.
Namely, the goal is to realize $\sum_i c_i U_i$ given a set of unitary gates $U_i$.
The coefficients $c_i$ can be encoded in a unitary gate $W$ serving as a measurement basis. 
Namely, on state $\sum_i |i\ket U_i |\psi\ket $,
the gate $W=[w_{ij}]$ maps the state to $|0\ket \sum_i w_{i0} U_i |\psi\ket$ 
plus other irrelevant terms.
The parameter $w_{i0}$ is proportional to $c_i$ with a proper normalization.
The post-selection on a multi-qubit state $|0\ket$ is required,
making it probabilistic. 
Such a probability can be boosted via the QAA algorithm. 

This algorithm can be used to construct gradient descent. 
For a Hamiltonian in the form $H=\sum_i c_i U_i$,
the algorithm realizes $H |\psi\ket$, which, according to Schr\"odinger equation,
is the derivative $i |\dot{\psi}\ket$. 
If there are varying parameters $\B \theta$, 
this can be used to find the optimal values of them. 
A common task is to find the ground state and energy of a model,
widely used in combinational optimization problems.


\subsection{Quantum simulation}

\begin{figure}[t!]
    \centering
    \includegraphics[width=0.35\textwidth]{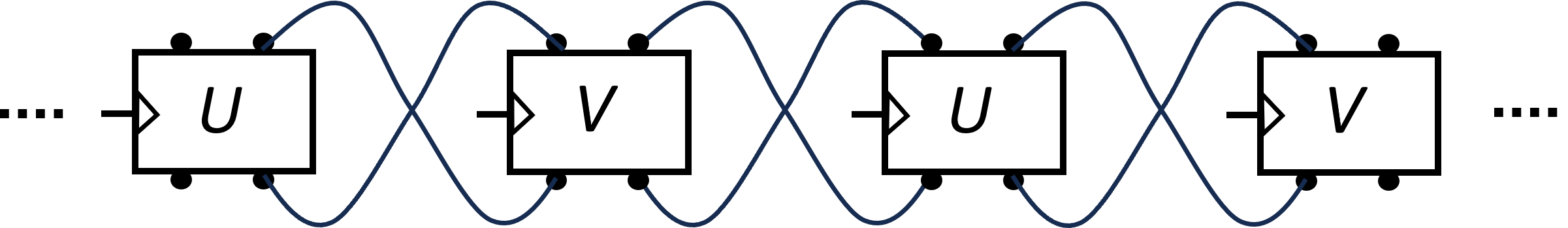}
    \caption{The chain of quantum transistors for quantum simulation. 
    Each wavy line between two edge modes represents ebits.  
    }
    \label{fig:qca}
\end{figure}

Quantum simulation of Hamiltonian evolution $e^{-itH}$ is one of the original motivation 
for quantum computing. 
A basic method is the Trotter-Suzuki product formula~\cite{NC00},
and here we analyze a primary setting. 
For a local Hamiltonian $H=\sum_n H_n$, the local terms $H_n$ can be grouped into 
commuting subsets, e.g., an even set and odd set for two-local interactions.
The circuit will be of a brickwork structure,
and a similar setting occurs in quantum cellular automata~\cite{Arr19}.
It contains iterative layers of gates $(\otimes_i U_i)(\otimes_j V_j)$
for $U$ and $V$ as the building blocks. 
The iteration can be realized by loops and reusable transistors,
illustrated in figure~\ref{fig:qca}. 
Referring to the original circuit, 
the horizontal direction is the space with the edge modes carrying the qubits,
while the time direction is `compactified' in the loops,
i.e., the measurement and refreshing sequence.

\section{Higher-level logical functions}

\subsection{Quantum programming}\label{sec:qp}

To construct large-scale integrated circuit,
programmability is a central feature.
It requires reconfigurable `wires' or connections among the circuit elements.
Notable examples include memory devices such as ROM, RAM,
and computing units such as FPGA~\cite{HH13}.
Note in literature ``quantum programming'' sometime is used to
refer to the design or compiling of quantum circuits,
while here our term has a unique meaning.

For the quantum case, both classical programming and quantum programming
have been analyzed, with the former refers to reconfigurable `wires' (i.e.,
free evolution between gates),
while the latter uses both reconfigurable `wires' and gates~\cite{W22_qvn}.
Namely, quantum programming employs superchannel~\cite{CDP08a} as the operation 
to change the function of a gate.
Given a channel $\Phi$, which can be represented as a unitary gate acting on a larger space,
the action of a superchannel $\hat{\C S}$ is 
\be \hat{\C S} (\Phi)(\rho)= \text{tr}_a V \; \Phi\; U (\rho\otimes |0\ket \bra 0|),  \ee
for a pre- unitary gate $U$ and a post unitary gate $V$, and a is an ancilla
with initial state $|0\ket$~\cite{WW23}. 
The channel $\Phi$, gates $U$ and $V$ can all be realized by transistors,
but one can also use hybrid circuits to realize quantum programming depending on the context.
Apparently, a sequence of superchannels can be connected together. 

Many algorithms we analyzed above are special types of quantum programming~\cite{W24rev}. 
The quantum control module maps $U$ to its controlled version $\wedge_U$.
The LCU algorithm maps a set $\{U_i\}$ to another one as a linear combination of them.
The QSVT maps a matrix $A=U\Sigma V$ to another one $B=U P(\Sigma) V$ for $P$
as a certain polynomial function of its singular value matrix $\Sigma$. 
It also naturally applies to the oracle setting by treating the input gates 
or channels as black boxes. 

A nontrivial component in integrated circuit is the address of a data cell.
Although allowing superposition is possible, 
here we use classical address,
similar to our choice of timing for synchronization in Sec.~\ref{sec:syn}.
Allowing superposition will generate entanglement between the address 
(or time) register and data register,
and this could achieve a certain tasks, 
as we will see in Sec.~\ref{sec:qst},
but also require a preparation stage and
a decoupling scheme to extract the data. 

Using classical space and time does not affect the universality and programmability 
of quantum circuits, and actually
there are always classical ingredients in a quantum algorithm or system. 
In the usual circuit model~\cite{NC00}, 
a quantum program is usually a unitary circuit $U$,
assuming standard basis initialization and readout.
It is stored as a classical program $[U]$ that describes the gates in it.
In quantum von Neumann architecture~\cite{W22_qvn},
the circuit $U$ can be stored as its Choi state $|U\ket$, 
and superchannels can be used to change its function and connect various gates.
On top of that, there is always a classical description of an algorithm, 
as a classical instruction set or program. 
All gates can be stored in transistors,
and in order to fetch a quantum program, 
the classical instructions are needed to execute the gates and measurements. 


\subsection{Quantum error correction circuits}\label{sec:qecc}

\begin{figure}[t!]
    \centering
    \includegraphics[width=0.35\textwidth]{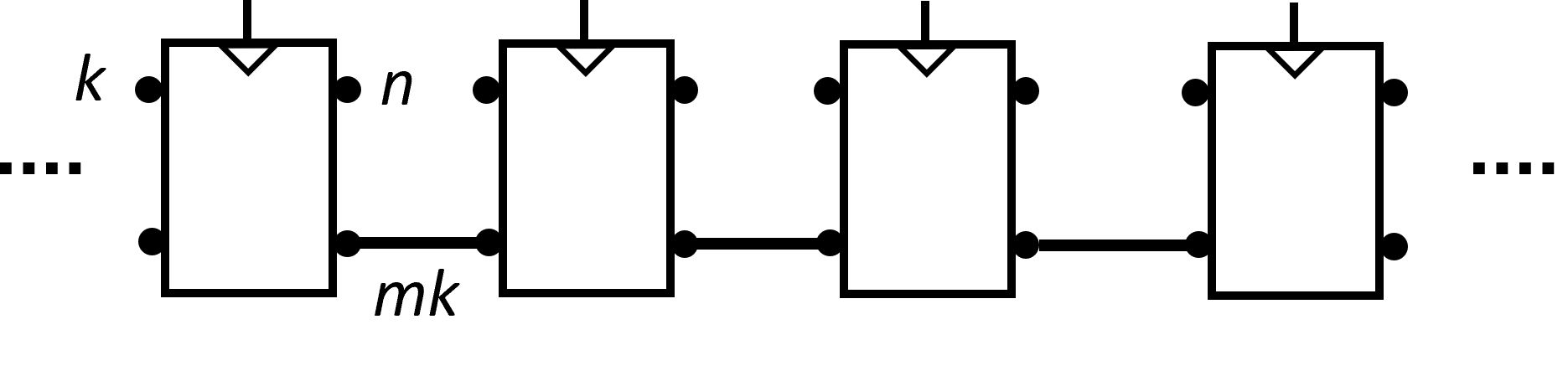}
    \caption{Quantum convolutional codes in the form of matrix-product states.
    The horizontal direction is time and each stage is a clock cycle.
    For each cycle, $k$ data qubits are taken as input and encoded into $n$ qubits,
    and $mk$ qubits as memory are feed forward to the next cycle.
    Here each transistor contains a $m$-stage quantum shift register. 
    }
    \label{fig:convcode}
\end{figure}

In this section, we show that QSCs can also be used for the study of quantum error correction codes.
As the classical case, there are in general two types of codes: 
\emph{block} codes and \emph{convolutional} codes~\cite{RL09,WLWL24}. 
The latter naturally deals with streaming data for real-time applications of low latency,
and it has a close relation with sequential circuits. 
In quantum computing, streaming data may apply to flying photons or flowing atoms,
so here we analyze the scheme to construct quantum convolutional codes.

A convolutional code is usually denoted as $(n,k,m)$ for one cycle,
with $k$ input bits encoded as $n$ output bits, 
and a $m$-stage shift register to store the $k$ bits from the previous cycle.
This structure carries over to the quantum case, shown in figure~\ref{fig:convcode}.
It is clear to see the overall structure is matrix-product state,
while here the modes for each `tensor' have different sizes ($k$ and $n$),
and the `bond' dimension is of size $mk$.
Each `tensor' is a hybrid circuit containing a $m$-stage quantum shift register
and a combinational  circuit.
Given $k$ input qubits, they are stored in the quantum shift register,
and together with the internal states of the register,
the state for $n$ output qubits and $mk$ memory qubits is generated 
by a unitary process. 
The output at a cycle would depend on the input at previous $m-1$ cycles,
with $m$ as a parameter for `correlation length'.
The time direction can be further compactified by looping the memory qubits
from the output modes back to the input modes. 

The connection between quantum convolutional codes with MPS 
has been studied~\cite{FP14},
and some stabilizer convolutional codes have been constructed~\cite{Wil09}. 
Due to the MPS form the codewords have limited entanglement, 
the code distance is rather small compared with some large-distance block codes,
and there are fast decoding algorithms such as the Viterbi algorithm~\cite{PP13}. 
However, its complexity increases exponentially with $m$.
When combined with block codes such as LDPC codes, 
convolutional LDPC codes can be defined 
which would require other types of decoding algorithms, 
such as belief propagation~\cite{RL09}. 
As qubits have short coherence time, 
some quantum LDPC codes can be converted into a convolutional one, 
so that input qubits can be generated at different times
to avoid idler error. 
This could be better than preparing all input qubits at once.




\subsection{Quantum space and time}\label{sec:qst}

In the above we have assumed classical space and time for some functions 
such as synchronization in Sec.~\ref{sec:syn} and memory device in Sec.~\ref{sec:qp}. 
On the contrary,
it is indeed feasible to 
generate superposition over spatial or temporal degrees of freedom. 
Such superposition have been widely studied, e.g., 
for photonic time-bin qubits and dual-rail qubits~\cite{KMN+07}, quantum walks~\cite{Ven12},
Feynman–Kitaev history state~\cite{KSV02}, 
and also QRAM~\cite{GLM08}. 
For convenience, we refer to such a register as a quantum \emph{spacetime} register, 
even though it may encode information exclusively in either spatial or temporal modes,
and can, in principle, be simulated using standard qubit registers without explicit spacetime superposition. 
Below we investigate a physical realization of a genuine quantum spacetime register using photonic systems.




Consider a set of stored quantum gates $\{U_i\}$ intended to act on a data register. 
A fundamental operation is to load these gates coherently via a spacetime control register
generating a state of the form
\be
|\Phi\rangle = 
\sum_{i=0}^{d_c-1} c_i \,|i\rangle  U_i |\psi\rangle,
\label{eq:qmux}
\ee 
where $|\psi\rangle$ is the input data state, 
$|i\rangle$ denotes a basis state of the control register, 
and $c_i$ are coefficients. 
States of this form can be realized by QMUX and 
arise in several settings: 
in DQC1, where measurement of the control register yields the trace of a unitary; 
in LCU, where 
post-selection on the control register implements an operator that can be non-unitary; 
in the Feynman–Kitaev method, where 
such a state encodes the entire history of a quantum circuit; 
and in QRAM 
it represents a parallel query over a database with items \(|D_i\rangle = U_i|\psi\rangle\).

\begin{figure}[t!]
    \centering
    \includegraphics[width=0.4\textwidth]{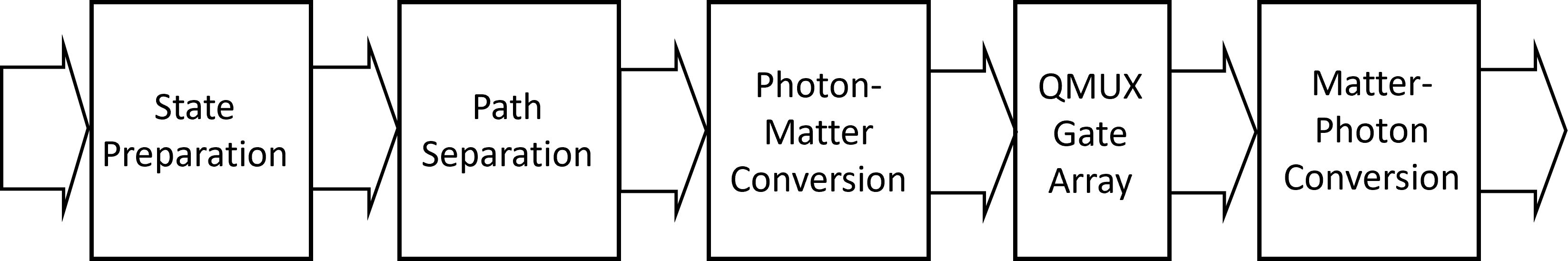}
    \caption{The illustration of using quantum spacetime register to realize 
    quantum multiplexer (QMUX). 
    The State Preparation yields 
    the initial data state $|\psi\rangle$ carried by 
    polarization of photons.
    The Path Separation yields the path state $\sum_i c_i |i\ket$ 
    and separates each path state $|i\ket$. 
    The Photon-Matter and Matter-Photon Conversion downloads or uploads 
    photon states.
    The QMUX gate-array contains a gate module $U_i$ placed on each path realizing 
    the parallel operation on $|\psi\rangle$.
    A Path Combination may be needed if further operation on the paths
    is required. 
    }
    \label{fig:qspacetime}
\end{figure}

Our architecture encodes the spacetime register in photons 
using two distinct photonic modes per photon: 
path for the control system and polarization for the target (data) system.
While others can also be used, such as orbital angular momentum,  
we focus on the polarization–path encoding for concreteness~\cite{KMN+07}. 
The basic task is to implement a controlled version \(\wedge_U\) of a given unitary \(U\),
and the QMUX in general. 
Prior work has addressed this for qubit systems~\cite{AFC14}; 
here we extend the construction to arbitrary dimensions. 
Consider a control register of dimension \(d_c = 2^k\) 
and a data register of dimension \(d_t = 2^m\), 
and a set of \(d_c\) distinct unitaries \(\{U_i\}\). 
In a photonic platform, 
this can be achieved without resorting to controlled-SWAP gates as discussed in Sec.~\ref{sec:qmux}:
photons can be separated into different spatial paths by cascade network of beam splitters (BS), 
and each path is equipped with the corresponding unitary \(U_i\) acting on the polarization-encoded data. 
A schematic is provided in the figure~\ref{fig:qspacetime}.
This scheme shall be combined with matter–photon conversion interfaces, e.g., 
via teleportation-based state transfer, 
to load or extract information between photonic and matter-based qubits. 
A product state of the photons is converted by the QMUX to the entangled state~(\ref{eq:qmux}).


Next we show how to apply the QMUX to other important tasks. 
The Feynman–Kitaev history states are used to study Hamiltonian complexity and 
adiabatic quantum computing~\cite{AL18}.
For a quantum circuit of depth \(T\), the history state is
\be
|\eta\rangle = \frac{1}{\sqrt{T+1}} \sum_{t=0}^{T} 
|t\rangle  |\psi_t\rangle,
\ee 
for $|\psi_t\rangle = U_t \cdots U_1 |\psi_0\rangle$.
We adopt a domain-wall encoding for the clock register: 
\(
|t\rangle = |0\rangle^{\otimes t} |1\rangle^{\otimes T-t}.
\)
The clock state 
\(
|\mathrm{clock}\rangle = \frac{1}{\sqrt{T+1}} \sum_{t=0}^{T} |t\ket
\)
is not a stabilizer state for \(T \geq 2\): 
its support contains \(T+1\) basis states 
(generally not a power of two), 
and its preparation requires non-Clifford rotations. 
It can be synthesized via a sequence of controlled rotations 
around Pauli $Y$ axis with 
\be
\theta_\ell = 2\cos^{-1}\!\Bigl(\frac{1}{\sqrt{T-\ell+2}}\Bigr), \qquad \ell = 0,1,\dots,T-1,
\ee
where the rotation on photon \(\ell\) is conditioned on the photon \(\ell-1\).
In our photonic setup, the clock register is encoded in path (domain-wall form) 
and the data register in the polarization of the same photons. 
Applying \(\wedge_{U_1}, \wedge_{U_2}, \dots, \wedge_{U_T}\) 
in parallel yields the state \(|\eta\rangle\).

In the QRAM setting, let data items be given by \(|D_i\rangle = U_i|\psi\rangle\) 
for a fixed initial state \(|\psi\rangle\). 
The goal is to perform the parallel query
\be
\sum_{i=0}^{d_c-1} c_i |i\rangle |\psi\rangle \;\longrightarrow\; \sum_{i=0}^{d_c-1} c_i |i\rangle |D_i\rangle.
\ee
Our QMUX architecture directly implements this transformation: 
the address register is encoded in the path of photons, 
and each \(U_i\) is placed in the corresponding spatial branch after the BS array. 
For the common case of a uniform superposition (\(c_i = 1/\sqrt{d_c}\)), 
the address state is straightforward to prepare.
The conventional bucket-brigade QRAM employs a binary tree of three-level routing nodes, 
achieving query complexity \(\mathcal{O}(\log d_c)\) with \(\mathcal{O}(d_c)\) physical nodes~\cite{GLM08}. 
Our approach trades the logarithmic query time for constant-time parallel selection, 
at the cost of requiring \(d_c\) distinct \(U_i\) networks. 
If each \(U_i\) is an arbitrary \(m\)-qubit unitary, 
the hardware overhead scales as \(\mathcal{O}(m^2 d_c)\). 
This can be advantageous when \(d_c\) is moderate and the \(U_i\) are structurally simple.

For the LCU algorithm, a further post operation on the path d.o.f. is also needed~\cite{Long06}.
This can be realized by a corresponding network of BS (and phase shifters)
to re-combine the paths together,
and usually a post-selection is needed.
Note the amplitudes $c_i$ in the initial state and the LCU form 
relate to each other but are not the same. 

Our design is modular and amenable to hybrid quantum hardware. 
An experimental challenge is the preparation of multiphoton entangled states, 
due to the inherently probabilistic nature of photonic entangling gates. 
We outline two scalable approaches. 
First, the cluster-state method enables universal state generation via adaptive measurements 
on a large photonic cluster state, 
which can be built from smaller entangled resources using fusion gates~\cite{BBB21}. 
Second, one may prepare the desired state deterministically on a matter-based platform 
(e.g., trapped ions or superconducting qubits) 
and then transfer it to photonic modes via matter–photon conversion interfaces. 


\section{Conclusion}

In this work, we defined quantum sequential circuits
and studied their usages to several classes of problems. 
These circuits are based on a new construction of quantum hardware
that we call quantum transistors. 
This serves as a hardware platform for realizing 
quantum von Neumann architecture~\cite{W22_qvn} 
and offers a new pathway toward scalable quantum information processing, thereby
complementing current qubit-based quantum processors. 

Our study makes the analog between classical and quantum computers 
more apparent. 
The general rules of modular and hierarchical design apply to both of them.
As in the classical case, 
the functionality of a sequential circuit can
be simulated by a combinational  circuit. 
However, the overhead may be significant. 
From a hardware perspective, different devices to carry bits 
or gates differ in many aspects in terms of their 
speed, energy cost, latency, and stability etc.
Therefore, introducing new type of circuit elements,
especially the primitive quantum transistors,
could help for the design and realization of large-scale 
integrated quantum chips. 

The quantum transistors defined in this work are based on
measurement and teleportation,
and in particular, along the research line of symmetry-protected
measurement-based quantum computing~\cite{ESB+12,NW15,SWP+17,ROW+19,RYA23}.
This makes them one-time and need to be re-settable,
and it remains to see 
if other viable schemes are possible.
Unitary evolution driven by Hamiltonian is desirable, 
e.g., adiabatic evolution~\cite{BFC13,WB15}, 
state transfer~\cite{AK25} or quantum walk~\cite{CGW13},
yet they must be controllable to enable the 
transistor’s on/off switching function, 
and modular as well as universal.

This work is also a hybrid of a few universal quantum computing models~\cite{W21_model},
which are often employed separately. 
Our work promotes for a pragmatic, integrated approach over an exclusive choice between existing models,
and suggests that there are still plenty of spaces to explore for advanced
quantum computing models or architectures, e.g., Refs.~\cite{YVC24,SCG+25,SCI+25,FRP+26}. 
Furthermore, it also leads to an intriguing question:
how to properly reconcile the qubit-based and transistor-based architectures? 
While classical computers offer a clear historical precedent, 
but for quantum computers this remains uncertain. 
Finally, our framework of sequential circuits may also be adapted 
to other applications of quantum information, 
such as feedback quantum control and communication.








\section{Acknowledgement}
This work has been funded by
the National Natural Science Foundation of China under Grants
12447101 and 12105343.

\end{spacing}


\bibliography{ext}{}
\bibliographystyle{elsarticle-num}

\end{document}